\documentclass[Afour,sageh,times]{sagej}
\usepackage{xurl}
\usepackage{moreverb,url}
\usepackage[colorlinks,bookmarksopen,bookmarksnumbered,citecolor=red,urlcolor=red]{hyperref}
\usepackage[utf8]{inputenc}   
\usepackage[T1]{fontenc}      
\usepackage{natbib}           
\usepackage{multirow}
\usepackage{float}            

\usepackage{array}        
\usepackage{longtable}
\usepackage{booktabs}
\usepackage{authblk}

\newcommand\BibTeX{{\rmfamily B\kern-.05em \textsc{i\kern-.025em b}\kern-.08em
T\kern-.1667em\lower.7ex\hbox{E}\kern-.125emX}}

\def\volumeyear{2025}

\usepackage{graphicx} 
\usepackage{amsmath,amssymb,amsfonts,amsthm}
\usepackage{todonotes}
\usepackage{thmtools} 
\usepackage{thm-restate}
\usepackage{float}
\usepackage{longtable}
\usepackage{booktabs}
\usepackage[section]{placeins}

\usepackage{hyperref}
\hypersetup{
 colorlinks=true, 
 linkcolor=red,  
 citecolor=teal, 
 urlcolor=red,    
 filecolor=teal 
}

\begin{document}

\runninghead{Operationalizing Justice}

\title{Operationalizing Justice: Towards the Development of a Principle Based Design Framework for Human Services AI}

\author[1]{Maria Y. Rodriguez}
\author[1]{Seventy Hall}
\author[4]{Pranav Sankhe}
\author[3]{Melanie Sage}
\author[2]{Huei-Yen Winnie Chen}
\author[1]{Atri Rudra}
\author[1]{Kenny Joseph}

\affil[1]{Department of AI and Society, University at Buffalo }
\affil[2]{Department of Industrial and Systems' Engineering, University at Buffalo}
\affil[3]{School of Social Work, University at Buffalo}
\affil[4]{Oak Ridge National Laboratory}


\email{myr2@buffalo.edu}

\begin{abstract}
Scholars investigating ethical AI, especially in high stakes settings like child welfare, have arguably been seeking ways to embed notions of justice into the design of these critical technologies.  These efforts often operationalize justice at the upper and lower bounds of its continuum,  defining it in terms of progressiveness or reform. Before characterizing the type of justice an AI tool should have baked in, we argue for a systematic discovery of how justice is executed by the recipient system: a method the Value Sensitive Design (VSD) framework terms Value Source analysis. The present work asks: how is justice operationalized within current child welfare administrative policy and what does it teach us about how to develop AI? We conduct a mixed-methods analysis of child welfare policy in the state of New York and find a range of functional definitions of justice (which we term principles). These principles reflect more nuanced understandings of justice across a spectrum of contexts: from established concepts like fairness and equity to less common foci like the proprietary rights of parents and children. Our work contributes to a deeper understanding of the interplay between AI and  policy, highlighting the importance of operationalized values in adjudicating our development of ethical design requirements for high stakes decision settings.
\end{abstract}





\maketitle

\section{Introduction}
Administrative policy - the rules and regulations that govern how a given system operates - can be understood in part as the institutionalization of the values and ideals a system wishes to hold stably over time \cite{shapiro_choice_1964}.  While not always understood through the lens of policy, the idea that technology institutionalizes values and ideals within a sociotechnical system is a critical line of inquiry \cite{selbst_fairness_2018}. Technology is itself a form of policy, whose values and ideals are informed by its creators, and in the context of machine learning, by the data used for training \cite{green_fair_2018,hoffman_discretion_2018,silva_algorithms_2018,suresh_framework_2019}. 

One question which arises is what values \emph{should} our technologies (and by extension our sociotechnical systems) institutionalize as policy? One notable effort to address this question is \emph{value sensitive design (VSD)}, emerging from Human-Computer Interaction (HCI) literature as a framework offering guidelines for the development of value-laden technology. More specifically, VSD offers 12 ethically centered values that tech likely should embody \cite{friedman_survey_2017, friedman_value_nodate}, while specifying the list to be at once non-trivial and non-exhaustive. VSD hints at the relationship between administrative policy and the values of technology in a method it calls value source analysis \cite{friedman_value_nodate}. \cite{le_dantec_values_2009} has argued that the methods of VSD may be furthered by considering the contextual values of a given technology prior to its deployment, a concept they term value discovery. However, value discovery is primarily aimed at the individual level rather than the system level \cite{le_dantec_values_2009, friedman_survey_2017}. 

The current paper argues administrative policy is an overlooked value source in the value sensitive design canon: administrative policy not only limits which values technologies should embody, but also \emph{predetermines how those values are operationalized within technological products}. More broadly, we note that within the HCI literature, while scholars have used VSD and other frameworks to develop fair, accountable, ethical, and/or justice-centered technology, and/or to critique the, e.g., fairness of existing technologies, there has been considerably less thought given to how new or proposed technologies may conflict with the existing policy landscape.

There are a variety of values, ideals and/or principles which technology could and perhaps should account for in its development, deployment, and iterations. However, a good deal of work has centered on values which pertain to ethical conduct, of both humans and the machines they create. For example, VSD defines human values as: 
\begin{quote}
    ...what is important to people in their lives, with a focus on ethics and morality \cite{friedman_value_nodate}
\end{quote}
No value is perhaps more debated in the child welfare and AI space than justice: what is the right thing to do \cite{sandel2007justice} in the face of (alleged) child abuse and neglect? Scholarship has examined whether and how actions in the child welfare system may or may not align with justice as fairness (for example \cite{dettlaff2023interrogating} and \citet{eubanks2018automating}), but little is known about what, if any, other conceptualizations of justice are described or implied within existing child welfare policy. Further, scholarship has not tied system sourced conceptualizations of values like justice with design principles for AI, particularly in high stakes settings.  The current manuscript addresses this gap by asking: how is justice operationalized within current child welfare administrative policy and what does it teach us about how to develop AI? 

Our work makes three contributions to the literature on AI and child welfare:
\begin{itemize}
    \item Conceptually, we provide a missing link between work on AI in child welfare with work on law and policy in child welfare through a value source analysis
    \item Empirically, we characterize the operationalization of justice within current (as of 2020) child welfare policy in New York state through a mixed-methods approach, and discuss how this emerging justice value space can help explain and/or better understand the current context of AI tools employed within child welfare, such as \cite{lanier2020preventing} 
    \item We provide the groundwork for a lens on AI in human services that understands administrative policy as a value source from which AI developers should acknowledge upper and lower bounds of technology design.
\end{itemize}

\section{Literature Review}

\subsection{A brief review of child welfare AI applications}
Over a dozen child welfare agencies used Artificial Intelligence (AI) in 2021 \cite{samant2021family}, with systems in place to help identify youth at risk of abuse \cite{Chouldechova_Benavides-Prado_Fialko_Vaithianathan_2018,Vaithianathan_Rouland_Putnam-Hornstein_2018}, to help select youth for social services interventions \cite{Saxenaetal2020}, and for various other purposes \cite{saxenaHumanCenteredReviewAlgorithms2020a}. Two systematic reviews of the use of predictive analytics in child welfare, one from a computer science perspective \cite{saxenaHumanCenteredReviewAlgorithms2020a}, and the other from a human services perspective \cite{hall2023systematic} explore this topic in more fine-grained detail. Additionally, various child welfare serving institutions have published reports on the potential challenges and opportunities posed by the use of predictive analytics in this sector  \cite{samant2021family,teixeira2017predictive,makingthemost,shroffPredictiveAnalyticsCity2017}. We refer the interested reader to these and provide only a brief summary of the most relevant work here.

At a high level, relevant work can be subdivided into three lines of effort. First, a broad array of work aims to better understand the decision-making process in modern child welfare organizations \cite{wurst2024context} and how algorithms, predictive analytics, and/or AI can play a role in it \cite{saxena2021framework,saxena2024algorithmic}.  Much of this work emphasizes a number of problematic uses of AI in this context \cite{keddellAlgorithmicJusticeChild2019}. \citet{lanier2020preventing}, for example, discuss Birth Match, a tool employed in New York and several other states (see for example, \cite{cohen_learning_nodate}) which aimed to identify high-risk infants \emph{at birth} given a mother who had their parental rights terminated at some past point. A number of others have studied racial biases embedded in the Allegheny Family Screening Tool \cite{Chouldechova_Benavides-Prado_Fialko_Vaithianathan_2018}, used to make decisions about whether or not to investigate a particular allegation of abuse or neglect. These works seek, implicitly or explicitly \cite{lanier2020preventing}, to characterize the values inherent in the development of specific AI tools. Second, some work has focused on the development of new tools--- AI-based and beyond--- that address existing challenges. Here, scholars have focused on technical considerations, e.g. in adding fairness constraints to existing algorithms \cite{Chouldechova_Benavides-Prado_Fialko_Vaithianathan_2018}; places where humans and AI might compliment each other \cite{chengHowChildWelfare2022,kawakamiImprovingHumanAIPartnerships2022}; ways to improve collaboration across different collections of human stakeholders \cite{stapletonImaginingNewFutures2022,kawakami2024situate}; as well as the potential for policy change \cite{du2022data}.  The present work compliments both sets by illustrating how the context a tool is deployed in offers a bounded space in which any given value may be expressed by any given tool. 

A third line of organization for work in the child welfare and AI space concerns proposed solutions for the myriad issues found in implementations of AI in child welfare to date. We characterize these as operationalizing justice at opposing ends of its spectrum. On one hand are works that see justice in a technical and perhaps reformist light \cite{Chouldechova_Benavides-Prado_Fialko_Vaithianathan_2018}, followed by the more sociotechnical vein \cite{kawakami2024situate,kawakamiImprovingHumanAIPartnerships2022} culminating in the sociopolitical \cite{du2022data}, which is oriented towards more progressive formulations of justice. Across all solutions, the goals are broadly the same: to help, rather than to hurt. Simultaneously, reformist and progressive AI solutions in child welfare can both cause unintended consequences: either reifying existing biases found in normative implementations of justice \cite{abdurahman2021calculating}, or removing services and assistance which some youth and families might genuinely benefit from in a more reformist implementation. It is likely there are undiscovered AI-based solutions that lie along the axes of reformist to progressive operationalziations of justice, and from net harm to net benefit for \emph{all} families and youth. 

In summary, work on child welfare AI tools has focused on the values embedded in AI within child welfare, and/or how to change them. However, there has been much less discussion on whether and how the values within AI solutions connect to the values of the given system it is being deployed in (i.e. law and policy). One critical exception is the work of \citet{abdurahman2021calculating}, who emphasizes that new federal child welfare policy surrounding data is friendly to the continued construction of predictive analytics at the expense of racial equity. 

\subsection{Child Welfare Law and Values}

Considerable work exists on laws and policies governing the use of AI. Beyond child welfare, such work has focused heavily on how the law has responded (or failed to respond) to the development of AI \cite{huaEffectiveEnforceabilityEU2023,paniguttiRoleExplainableAI2023,valentineImpoverishedAlgorithmsMisguided2019,jonesIroniesAutomationLaw2015} and predictive analytics more broadly \cite{abdurahmanCalculatingSoulsBlack2021}, how new laws and policies on AI are implemented in practice \cite{lawrenceBureaucraticChallengeAI2023,pistilliStrongerTogetherArticulation2023} and on how concepts of fairness in AI relate to discrimination law \cite{royMultidimensionalDiscriminationLaw2023,weertsAlgorithmicUnfairnessLens2023,calviEnhancingAIFairness2023}.  
However, such work sidesteps the reality that \emph{the law itself is laden with values which shape the kinds of AI we pursue}.  Put another way, while many have worked to produce methods for better implementing AI in high-risk decision-making systems, there has been limited consideration of the rules of those systems as they are now - the values they espouse, the constraints they impose, and the outputs they require. 

In contrast, a number of scholars in child welfare have examined policy and its implications \cite{flukeDifferentialResponseChildren2019,connellyStateChildWelfare2014,gainsboroughScandalsLawsuitsPolitics2009,dayStateChildWelfare2022,rebbe_state-level_2024}. 
Perhaps most relevant to the present work in this vein is the development of the SCAN database \cite{weigensberg2020scan}. Released in 2020, the SCAN database presents a rich array of measures that characterize how each state defines child abuse and child neglect, as well as a number of other factors that are associated with child welfare policy. The SCAN database has been used to study the impact of policy on various outcomes in child welfare, including reunification \cite{labrenz_state-level_2023}, racial disparities in child maltreatment \cite{labrenz_state_2023}, and rates of allegation substantiation \cite{labrenz_state_2023-1}. The present work compliments the efforts of the SCAN database by 1) focusing on implicit and/or explicit operationalizations of justice, rather than definitions of abuse and neglect, and 2) focusing on the relationship between these values and AI/predictive analytics.

\subsection{Value Sensitive Design and Policy Analysis}

\citet{friedman_survey_2017} defines value sensitive design as a theoretically driven framework which seeks to "account for human values in a principled and systematic manner throughout the design process" (pg. 1). The authors survey 14 specific design methods for the application or uncovering of values in the tech design process. Importantly, we argue  \citet{friedman_survey_2017} defines values in a deductive way, offering methods and constructs to identify what might be important to specific people in the design process. Recent scholarship on VSD has underscored this point, noting that applications of the theory and its resultant methods often serve to evaluate the values of a technology after it has been deployed in a given system (what \citet{le_dantec_values_2009} calls ``ex post facto value analysis''), as opposed to examining system values prior to technological deployment.  Prior work on predictive analytics in child welfare offers a similar perspective \cite{rodriguez2019bridging}: value sensitive design may be well served by examining \emph{how values are enacted in a given sociotechnical system before the deployment of technology. In this work, we argue the a priori guardrails of policy limit the technology that can be developed in the first place}. 

Similar to \citet{le_dantec_values_2009}, we seek to ``discover'' how the child welfare system operationalizes a specific value, Justice, rather than compel it to our expectations. The aim of undertaking the investigation of justice in this way (i.e. at a system level) is to understand the ways in which technology can meet the system where it is. In social work practice, of which child welfare is central, this approach is called a strengths based approach \cite{solomon1987empowerment}. Importantly, \citet{friedman_survey_2017} notes that value source analysis is a methodology within the VSD canon. However, other than the case found in the same, we have not found other examples of this method in the literature. 

In this manuscript, we undertake a value source analysis of the child welfare system, using New York State as a case study. The justification for our  case study of New York is practical. As of 2022, there were a little over 14,000 youth in foster care in the state of NY (according to \citet{acf_afcars_2022}), making it approximately the 4th largest child welfare jurisdiction in the United States. Policy diffusion is a public administration theory which describes the ways in which policy ideas, tools, indeed policies themselves, move across governmental bodies (i.e. states, cities, townships, etc.).  One of these mechanisms is learning from early adopters \cite{shipan_mechanisms_2008}, in which states with lesser resources learn from the experiences of more innovative state actors with the resources and scale necessary to pilot untested ideas \cite{walker_diffusion_1969, mintrom_policy_1997, gray_innovation_1973}. New York is a known policy innovator, and research has demonstrated that its policies tend to influence federal policy (see, for example, \cite{giladi_effect_2014}. 

\section{Data and Methods}
The present work is part of a larger project that explores the content of child welfare policy in the United States. For the larger project, we began by collecting all child welfare relevant policy in the United States as of 2020. Then, using  the computational grounded theory approach described in \citet{rodriguez_computational_2020}, we conducted an initial round of qualitative coding on New York state. The present work centers on how justice is operationalized within New York state child welfare policy.

After data collection, following \citet{rodriguez_computational_2020}, we first applied a computational model (a topic model, specifically) to the full set of text data to assist in a first pass of qualitative coding. We analyzed the output of this topic model and used it to seed an initial qualitative code book. We then selected 379 policies from New York state and examine these for two additional rounds of finer grained qualitative coding. In the second pass, we refined the themes generated by the topic model. In the third pass,  we focused on a subset of the policy (detailed below) to identify the justice operationalizations that are encapsulated in these rules. Finally, we compliment our third round of codes using a computational method---similarity queries using sentence embeddings---that helped us to surface additional policy excerpts relevant to the themes we identified in our third round of coding. In this 4th round, two authors reviewed all sentence similarity output and mutually selected all relevant excerpts offered herein.  In total, the current analysis is the summation of 1 computational round of topic modeling, 2 qualitative coding rounds, followed by a 4th computational round of sentence similarity analysis. 

Qualitative researchers use the term "thick description" to define analysis which contextualizes the acts or documents that are under consideration. \cite{ponterotto_brief_2015} notes 'thick description' facilitates 'thick' interpretation, allowing researchers to bring "readers to an understanding of the social actions being reported upon" (pg. 542).  We use a case study approach in order to employ thick description towards our long term goal noted above: a framework that uses administrative policy as a value source to identify the contextual guardrails of AI in human service systems. Further, as a result of a meticulously documented process, we argue our computationally book-ended case study method can be replicated with other states.

In following sections, we briefly detail the steps taken in the larger project?: how we identified and collected child welfare policy, and how we conducted our initial coding on New York state policy. We then describe how we drew from this larger project to focus on the research questions relevant to the present work. Full details are provided in the appendix.

\subsection{Collecting Child Welfare Policy in the United States}

We consulted the \emph{Child Welfare Information Gateway}’s (CWIG) State Statutes search database as a guide to assist us in identifying state statutes and regulations relevant to child welfare \cite{cwig}. CWIG is an electronic resource center operating under the auspices of the U.S. Children’s Bureau that provides the public with access to information about child welfare policy, practice, and statistics. Entering a query into CWIG’s database generates a summary of a given state’s policies and regulations regarding child abuse and neglect, the child welfare system, and/or adoption. Users can filter their searches by selecting specific topics under each of these three categories; for example, a user interested in learning more about state policy on kinship care might select ‘kinship guardianship as a permanency option’ under the ‘child welfare’ search category. For the purposes of this project, we selected all topics listed under ``child abuse and neglect'' and ``child welfare,'' excluding adoption-specific statutes from our search results. This search was performed separately for each of the 50 states, Washington, D.C., and Puerto Rico. 

Statute citations were pulled from the generated CWIG reports and entered into a Google spreadsheet for future use. To ensure consistency in formatting within documents and uniformity in data collection procedures across states, we primarily relied on a single source, Casetext.com,\footnote{https://casetext.com/} to access PDF copies of statutes and regulations. Whenever we were unable to find a statute on Casetext, we referred to state government websites. Additional details on data collection, as well as some descriptive statistics of the full dataset, can be found in the appendix.

\subsection{Initial Coding of New York State Policy}

Before beginning our analysis, policy documents were split into excerpts at the sentence-level. Following \citet{rodriguez_computational_2020}, we began our analysis by applying a topic model to the set of all child welfare policy in order to get a broad sense of potential themes of interest to qualitative coding. At a high level, a topic model is a probabilistic model of text that assumes documents within a corpus can be represented as a small number of ``topics,'' where a topic is a distribution over words in a pre-specified vocabulary. Specifically, we use a structural topic model (STM) \cite{robertsStructuralTopicModels2014} for this task, as we find that it helps to address language differences across states. In the appendix, we detail the application of the topic model in more detail, as well as providing results from it. 

Following our use of topic modeling outputs to seed an initial codebook, the research team took the portion of the dataset pertaining to New York State and engaged in one round of qualitative coding.  To do so, we used thematic qualitative analysis \cite{braun_using_2006}. Each document was assigned 2 coders - one substantive expert, and one novice coder. To perform coding, the coding team met weekly during the summer of 2021 from June-August. During the coding meetings, the team discussed the codes they applied during the prior week, asked and answered questions concerning content and code application, and resolved discrepancies between and amongst code categories. The authors then took the codes that were developed and refined them as a team, to reduce redundancy, ensure clarity in coded themes, and define/refine each theme identified. Each code was reviewed by each of the substantive coders from the previous coding round. This process resulted in 481 substantive themes, with 282 identified nuanced themes within these. 

\subsection{Mixed Methods Analysis to Surface Principles of Justice}

Finally, we analyze a subset of New York state policy relevant to our research questions. To do so, we draw excerpts annotated with at least one of 15 substantive themes---from the 282 nuanced themes identified in our initial analysis described above---which we agreed would provide the strongest insights into how justice is operationalized. Themes selected were those that pertained to the \emph{elucidation of rules} within the child protective services process. The names of these themes (i.e. which we identified in our initial analysis) are depicted in Table~\ref{tab:substantive_themes} in the appendix. 

Narrowing our analysis to excerpts coded with at least one of these themes, we arrived at a final set of n=5,746 excerpts. Our analysis proceeded in a series of steps. First, we used an inductive process of qualitative coding, wherein we allowed the excerpt text to define the value itself \cite{saldana_coding_2009}. More specifically, over the course of six months in 2024, the lead author of the paper conducted an initial inductive coding to surface potential justice realizations for inclusion in the study. Following this initial inductive process, over the course of three months in 2024, the lead author and another senior author discussed and refined these codes through further qualitative coding.

Finally, we conducted a computational 4th round of analysis, to provide further refinement and validation of our qualitatively derived codes. To do so, we first draw on approaches that use (projections of) embeddings to surface text that is relevant to a concept of interest. While this approach has largely been used with word embeddings \cite{kozlowskiGeometryCultureAnalyzing2019,gargWordEmbeddingsQuantify2018}, recent work has extended this approach to embeddings within short documents \cite{senInsiderStoriesAnalyzing2023}. For each principle identified our initial inductive coding,we construct a set of concepts representative of that principle. Then, using a sentence embedding model (SimCSE \cite{gao2021simcse}), embed these sentences and all policy excerpts. Then, for each principle, we identify the 20 most similar excerpt embeddings (using cosine similarity) to that principle. Finally, we conduct another round of qualitative analysis on these excerpts, using them to surface examples of each principle and/or to refine our final definitions of the principles. Sentences used for each principle are provided in Table~\ref{tab:simcse} in the appendix, the top 20 excerpts for each will be provided in the data release for this paper.

\subsection{Positionality Statement}

\textbf{A positionality statement will be included in the final article as standard practice notes it should not be included in an anonymous submission.}

\section{Results}

We identified thirteen operationalizaitons of justice (herein after referred to as Justice  principles) reflected in New York State child welfare policy. Table~\ref{tab:values} offers the 13 justice principles we identified along with a brief description, and  Table~\ref{tab:principles_and_excerpts} provides example policy excerpts for each principle along with the NYS administrative rule from which they originate; relevant excerpts from the table are also provided in the text. In this section, we address our primary research question, providing a detailed description of the principles and their intersections with how AI could be developed within child welfare (which we term AI guidelines). We summarize the principles using four groupings: defining justice, communicating justly, intangible justice, acting justly. 

\begin{table*}[t]
  \centering
  \footnotesize
  \begin{tabular}{p{3cm}p{11cm}}
    \toprule
    \textbf{Principle}  & \textbf{Definition} \\ 
    \midrule
    Transparency & A rule that is intended to state what must be made public or restrictions on what can be made public \\
    Well-being & A rule pertaining to (often, the child's) health or physical or emotional well-being \\
    Dignified experience & A rule pertaining to the paramountness of life \\
    Autonomy (of children) & A rule that emphasizes that an actor has freedom to make choices \\
    Proprietary Rights of Family Relations & A rule that describes or alludes to proprietariness between family members \\
    Liberty of Faith & A rule pertaining to the importance of religion in decisions being made \\
    Relationships of Care & A rule that reifies consanguinity – describing or defining how people within a CW case relate to each other by blood, birth, or association \\
    Protecting from exposure to harm & A rule pertaining to the goal of avoiding harm at any (or at a particular) cost \\
    Standards for action & A rule that tries not to pick a side and instead aims to adjudicate between all sides \\
    Fairness\textbf{*} & A rule that emphasizes decisions should be made with suitability and appropriateness to the situation, taking context into consideration \\
    Accountability\textbf{*} & A rule that clearly states who is responsible for something and what happens if they fail in that responsibility \\
    Equity\textbf{*} & A rule that acknowledges that something is contextual and that we have to think about that context when making decisions \\
    \bottomrule
  \end{tabular}
  \caption{The 13 justice principles (left column) we identified on our third round of qualitative coding and their definitions. * = Fairness, Accountability, and Equity}
  \label{tab:values}
\end{table*}

\clearpage
\onecolumn
\begin{longtable}{@{}p{0.15\textwidth} p{0.6\textwidth} p{0.15\textwidth}@{}}
\caption{Identified justice principle, exemplary excerpts, and the origins of the New York State administrative policy title the excerpts originate from}
\label{tab:principles_and_excerpts} \\

\toprule
\textbf{Principle} & \textbf{Excerpt} & \textbf{Policy} \\
\midrule
\endfirsthead

\multicolumn{3}{@{}l}{\small\textit{Table \thetable\ (continued)}}\\
\toprule
\textbf{Principle} & \textbf{Excerpt} & \textbf{Policy} \\
\midrule
\endhead

\bottomrule
\endfoot

\bottomrule
\endlastfoot

Transparency
  & Such notice shall be printed in both Spanish and English and contain in conspicuous print and in plain language the information set forth in this paragraph. \newline\newline  
    Intake procedures shall be clearly delineated in the documentation file. At a minimum, such documentation file shall include…  
  & Section 1055 \newline\newline NY 432.2 \\
\midrule

Well-being
  & …a comfortable, private setting that is both physically and psychologically safe for children. \newline\newline  
    …provide a family atmosphere of acceptance, kindness and understanding and endeavor to give each child the support, attention and recognition that facilitates adjustment to the home and that promotes the child’s normal development  
  & Section 423-A \newline NY 443.3 \\
\midrule

Protecting from Exposure to Harm
  & forensic interviews to be conducted in a manner which is neutral and fact-finding and coordinated to avoid duplicative interviewing. \newline\newline  
    The physical space, construction and maintenance of the foster home and premises must be in good repair and kept in a sufficiently clean and sanitary condition so that the physical well-being as well as a reasonable degree of physical comfort is assured the members of the foster family  
  & Section 423-A \newline\newline NY 443.3 \\
\midrule

Ethics
  & Caregivers shall apply the reasonable and prudent parent standard when deciding whether or not to allow a child in foster care to participate in age- or developmentally-appropriate extracurricular, enrichment, cultural, or social activities. \newline\newline  
    This article is designed to establish procedures to help protect children from injury or mistreatment and to help safeguard their physical, mental, and emotional well-being. It is designed to provide a due process of law for determining when the state, through its family court, may intervene against the wishes of a parent on behalf of a child so that his needs are properly met.  
  & Section 383-A \newline\newline\newline\newline Section 1011 \\
\midrule

Right to Life
  & Provided, however, that nothing in this subdivision shall be construed to limit the ability of a child to consent to his or her own medical care as may be otherwise provided by law. \newline\newline  
    Health supervision, medical and dental care must be provided to each youth in accordance with section 441.22 of this Title.  
  & NY 661 \newline\newline NY 449.4 \\
\midrule

Proprietary Rights of Familial Relations
  & …that the parent or guardian has the right, prior to signing the instrument transferring the care and custody of the child to an authorized agency, to legal representation of the parent’s own choosing. \newline\newline  
    Nothing in such order shall preclude any party from exercising its rights under this article or any other provision of law relating to the return of the care and custody of the child by a social services official to the parent, parents or guardian.  
  & Section 384-A \newline\newline Section 1015-A \\
\midrule

Relationships of Care
  & …the emotional relationship of the siblings to each other. \newline\newline  
    Indian tribe shall mean any tribe, band, nation or other organized group or community of Indians recognized as eligible for the services provided to Indians by the Secretary of the Department of the Interior or by the State of New York or by any other state because of their status as Indians.  
  & NY 431.10 \newline\newline NY 431.18 \\
\midrule

Autonomy of Children
  & Provided, however, that nothing in this subdivision shall be construed to limit the ability of a child to consent to his or her own medical care as may be otherwise provided by law. \newline\newline  
    Nothing in this subdivision mandates the participation of a child in the status of trial discharge or supervision. Such participation is contingent upon the consent of the child.  
  & NY 1706 \newline\newline NY 430.12 \\
\midrule

Liberty of Faith
  & …recognize and respect the religious wishes of the natural parents of children in care and endeavor to protect and preserve their religious faith. \newline\newline  
    In appointing guardians of children, and in granting orders of adoption of children, the court shall, when practicable, appoint as such guardians, and give custody through adoption, only to a person or persons of the same religious faith as that of the child.  
  & NY 443.3 \newline Section 373 \\
\midrule

Accountability
  & That the failure of the parent or guardian to meet the obligations listed in subparagraph (v) could be the basis for a court proceeding for the commitment of the guardianship and custody of the child to an authorized agency thereby terminating parental rights. \newline\newline  
    Failure on the part of staff to acknowledge the code of conduct can result in disciplinary action including termination, consistent with appropriate collective bargaining agreements.  
  & Section 384-A \newline\newline NY 449.4 \\
\midrule

Fairness
  & The type and level of a foster care placement for a particular child shall be considered appropriate for the purposes of this section if the standard for continuity in the child’s environment and the standards for appropriate level of placement, as set forth in subdivisions (c) and (d) of this section, are met. \newline\newline  
    For the purposes of this analysis, the standards for determining whether a placement is unavailable must conform to the prevailing social and cultural standards of the Indian community in which the Indian child’s parent or extended family resides or with which the Indian child’s parent or extended family members maintain social and cultural ties.  
  & NY 430.11 \newline\newline NY 431.18 \\
\midrule

Equity
  & An Indian custodian may demonstrate that he or she is an Indian custodian by looking to tribal law or tribal custom or State law.  
  & NY 431.18 \\
\end{longtable}

\clearpage
\twocolumn

\subsection{Defining Justice}

Several scholars have taken great pains to define fairness \cite{mitchellAlgorithmicFairnessChoices2021}, accountability \cite{brownAlgorithmicAccountabilityPublic2019}, and equity \cite{katell2020toward}. Despite taking an inductive approach that need not have surfaced these principles, we find that NYS child welfare policy also operationalizes these concepts as principles of justice. However, the policy documents reviewed here often do not use the terms as explicitly, rather providing a much more contextualized depiction of these notions. \textbf{Fairness} within NYS child welfare policy we find  adjudicates between what the scholarly literature terms fairness and equity and \emph{the needs of a child found to be in unsafe conditions}. 
\begin{quote}
    The type and level of a foster care placement for a particular child shall be considered appropriate for the purposes of this section if the standard for continuity in the child's environment and the standards for appropriate level of placement, as set forth in subdivisions (c) and (d) of this section, are met. (NY 430.11)
\end{quote}
\textbf{Accountability} in NYS child welfare policy recognizes the child welfare system to be 1) a mechanism to hold caregivers responsible for child-rearing practices and outcomes, and 2) a duty-bound entity that must continuously expose its decision-making processes for scrutiny. 
\begin{quote}
    That the failure of the parent or guardian to meet the obligations listed in subparagraph (v) could be the basis for a court proceeding for the commitment of the guardianship and custody of the child to an authorized agency thereby terminating parental rights (Section 384-A) 
\end{quote}
\begin{quote}
    Failure on the part of staff to acknowledge the code of conduct can result in disciplinary action including termination, consistent with appropriate collective bargaining agreements (NY 449.4)
\end{quote}

Finally, what we call\textbf{addressing inequity} recognizes that some decisions made by child welfare system actors must be delegated to more localized authorities, especially in cases where some kind of systemic prejudice might exist. 
\begin{quote}
    An Indian custodian may demonstrate that he or she is an Indian custodian by looking to tribal law or tribal custom or State law. (NY 431.18)
\end{quote}

Perhaps more than anywhere else, then, our findings on these three principles underscore the importance of understanding policy as a mechanism that constrains both potential and realized AI development in child welfare. While the fairness in ML community continues to develop novel and important work around better definitions and operationalizations of these concepts, the literature shows child welfare practitioners hold policy to be paramount in their decision-making \cite{wurst2024context},  thus we should expect that AI tools developed with novel interpretations of fairness, accountability, and/or equity, even if they improve upon definitions within policy, are likely to be sidelined when actual decisions are being made. More concretely, the first AI guideline stemming from the current work is : \emph{AI child welfare tools wishing to optimize for justice principles should define the "best outcome" of any technological intervention or tool as part of a process whereby that child's primary, safe relationships can be nurtured and strengthened.} 

\subsection{Communicating Justly}
In contrast to the above three principles, we find the operationalization of the justice principle \textbf{transparency} in NYS child welfare policy provides a more contextualized yet broadly consistent definition in comparison to work in the fairness in ML literature. Specifically, we find that certain administrative rules underscore the importance of documentation and record keeping practices that ascertain the facts of a case as well as the flow of decisions as a result of those uncovered facts. We also find rules which pertain to transparency attempt to stabilize the standards of communication between child welfare involved and child welfare employed parties. 
\begin{quote}
    Such notice shall be printed in both Spanish and English and contain in conspicuous print and in plain language the information set forth in this paragraph (Section 1055) 
\end{quote}
Transparency in the child welfare system thus requires ascertaining that all parties receive the information they are entitled to regardless of the potential barriers that might preclude both parties from communicating effectively.  This principle also attempts to establish a threshold for the minimum amount of information that should be recorded, recognizing that such information should be understandable to most people. AI applications in the child welfare context that wish to follow this principle of justice in context would thus, much like the fairness in ML community has argued for in the past \cite{saxena2024algorithmic}, \emph{ensure that all the inputs and outputs of their tools have public facing documents that are understandable to all involved.}

In addition to the above principles of justice which the fairness in ML community has taken care to define, the specific context of child welfare also contains a number of distinct principles of justice. We review those in the remainder of this section.

\subsection{Intangible Justice}
The principle of \textbf{well being} sets the standard for what constitutes the outcomes of good care for children in the child welfare system, as well as describing \emph{the values that follow from it}. 
\begin{quote}
    ...provide a family atmosphere of acceptance, kindness and understanding and endeavor to give each child the support, attention and recognition that facilitates adjustment to the home and that promotes the child's normal development (NY 443.3) 
\end{quote}
This principle underscores how \emph{the quality of a decision is as important as the speed or preciseness of that decision}. Foster homes are the last resort for both the system employed and the system involved: if a child must be removed from their home, the above articulates the standard must be the home a child is taken to will be safer than the home they came from. To state this guideline broadly: \emph{the outcome of any decision (AI based or not) should leave the subject better than they were before the decision}. Many prior efforts in AI and child welfare have sought to assess the immediate risk of a child who was the subject of a child abuse allegation \cite{teixeira2017predictive, hall_systematic_2023, saxena_human-centered_2020}. However, such efforts do not account for whether a child will be safer when removed than where they are. This points to ways in which policy dictates new forms of evaluation that are critical for success. Moreover, the principle of well being suggests that such evaluations must be based on characteristics such as ``acceptance'' and ``understanding'' which are not easily quantified, and thus that do not lend themselves cleanly to the current practices of AI model evaluation in human service settings.

The principle of well-being is complemented by five other principles which similarly characterize  intangible principles of justice which are nonetheless reflected in NYS child welfare policy. The principle \textbf{dignified experience} asserts that children are non-trivial actors in child welfare cases, and must be afforded the ability to govern their own bodies. 
\begin{quote}
    Provided, however, that nothing in this subdivision shall be construed to limit the ability of a child to consent to his or her own medical care as may be otherwise provided by law. (NY 661) 
\end{quote}
The principle \textbf{autonomy of children} takes the above further by emphasizing that children who can speak can offer or decline consent to actions initiated by the state (and presumably other parties). 
\begin{quote}
    Nothing in this subdivision mandates the participation of a child in the status of trial discharge or supervision. Such participation is contingent upon the consent of the child. (NY 430.12) 
\end{quote}
The principle \textbf{proprietary rights of familial relations} holds the converse emphasis: though biological or familial caretakers find themselves under the auspices of child welfare proceedings, it does not lessen the belonging-ness of their children to them.   
\begin{quote}
      Nothing in such order shall preclude any party from exercising its rights under this article or any other provision of law relating to the return of the care and custody of the child by a social services official to the parent, parents or guardian. (Section 1015-A) 
 \end{quote}
Belonging-ness is further refined in the principle \textbf{liberty of faith}: 
\begin{quote}
    ...recognize and respect the religious wishes of the natural parents of children in care and endeavor to protect and preserve their religious faith (NY 443.3) 
\end{quote}
Finally, the principle \textbf{relationships of care} emphasizes the centrality of larger kinship relationships, according to child welfare policy, to a child's ability to be: 
\begin{quote}
    Indian tribe shall mean any tribe, band, nation or other organized group or community of Indians recognized as eligible for the services provided to Indians by the Secretary of the Department of the Interior or by the State of New York or by any other state because of their status as Indians. (NY 431.18)
\end{quote}

AI tools developed wishing to safeguard this group of principles would do so in at least two concrete ways. First, \emph{rows of data should be fundamentally understood as real independent actors whose future is not dictated by their past, but by their present} \cite{lanier_preventing_2020, rodriguez2019bridging}. Moreover, the critical elements of the present which are necessary for decision-making are largely qualitative in nature. 
 As such, in line with developments in child welfare policy over the past decade (\cite{noauthor_separate_2023} ), it is critical to acknowledge AI tools must be sensitive to the core need of relationship building and relational structures for each case, particularly between the caseworker and the youth and family \cite{curryIfYouCan2019, stalker_child_2007, sinai-glazer_essentials_2020}. Second, \emph{AI tools should look to relevant state/local policy to identify the core dimensions of evaluation, and where those dimensions are qualitative, interdisciplinary work is critical to determine how best to approach evaluation}.  

\subsection{Acting Justly}
The excerpts comprising the principle \textbf{protecting from exposure to harm} signal a key child welfare system perspective: the child welfare system assumes children's' interactions with it are, by definition, harmful. Having unknown adults ask questions about a child's family, how they treat them and, potentially, about their body, is jarring at the very least and re-traumatizing at worst.  NYS policy takes care to set the standard for how those kinds of interactions should occur:
\begin{quote}
    ...forensic interviews to be conducted in a manner which is neutral and fact-finding and coordinated to avoid duplicative interviewing (Section 423-A) 
\end{quote}
Further, policy stipulates that children in foster care require particular protection, in so far as the foster care environment they find themselves in (randomly) at minimum \emph{should not cause more harm.}  
\begin{quote}
    The physical space, construction and maintenance of the foster home and premises must be in good repair and kept in a sufficiently clean and sanitary condition so that the physical well-being as well as a reasonable degree of physical comfort is assured the members of the foster family (NY 443.3) 
\end{quote}

This principle is one of the most relevant to the development of AI because it is a core imperative of the system and predisposes it to type II errors \cite{rodriguez2019bridging}. The knowledge that exposure to the child welfare system is at minimum jarring to a child \emph{requires that the design of an AI tool scrutinizes the real-world outcome of the tool before the tools' quantitative strengths}. For example, \citet{lanier_preventing_2020} deeply engage with the contextual factors required to ethically implement an AI tool known as BirthMatch (implemented in 3 states at the time of that manuscript, including New York). The tool flags a newborn who is born to someone who has either 1) had their parental rights terminated previously or 2) has a child currently in foster care. Ostensibly, this requires that the name of a person who has just given birth be run through a model while they are recovering from labor in the hospital.  \citet{lanier_preventing_2020} note implementation of this tool would have far-reaching consequences: BirthMatch changes the life course for two generations (pg. 9), introduces at least one adverse childhood experience (ACE) \cite{felitti1998relationship} to a newborn, while basing its deployment on the need to prevent a harm that occurs in less than 1\%\ of high risk cases (pg. 9). That is, \emph{the principle of \textbf{protection from exposure to harm} is so central to how the child welfare system operates that any AI tool implemented within it needs to temper the tendency towards type II errors strongly while accepting the mandate to protect those whom are among the most vulnerable}. 

The principle \textbf{standards for action} reifies the centrality of justice in orienting the system towards actions that are a last resort in the face of due process supported evidence. 
\begin{quote}
    This article is designed to establish procedures to help protect children from injury or mistreatment and to help safeguard their physical, mental, and emotional well-being. It is designed to provide a due process of law for determining when the state, through its family court, may intervene against the wishes of a parent on behalf of a child so that his needs are properly met. (Section 1011) 
\end{quote}
This suggests an action standard cognizant of the tendency towards type II errors which attempts to mitigate that tendency by charging itself with the burden of proof. AI tools employed in child welfare should adopt this same principle, especially in a decision-making/augmenting context: \emph{the burden of proving the output of a tool as the least harmful course of action rests on the side of the tool developer.} The principle of "transparency" noted at the start of this section further requires the documentation of that proof to be made available publicly and in plain language to all parties at all times, regardless of the communication ability of any party. 

In summary, we find AI tools wishing to optimize for justice principles in human service contexts should:
\begin{enumerate}
    \item define the "best outcome" of any technological intervention or tool as part of a process whereby the subject's primary, safe relationships can be nurtured and strengthened
    \item ensure all inputs and outputs of the tools have public facing documents that are understandable to all involved
    \item ensure the quality of a decision is as important as the speed or preciseness of that decision: the outcome of any decision (AI based or not) should leave the subject better than they were before the decision
    \item ensure rows of data are fundamentally understood as real independent actors whose future is not dictated by their past, but by their present
    \item look to relevant state/local policy to identify the core dimensions of evaluation, and where those dimensions are qualitative, interdisciplinary work is critical to determine how best to approach evaluation
    \item scrutinizes the real-world outcome of an AI based tool before the tools' quantitative strengths
    \item temper the tendency of human services towards type II errors while accepting the mandate to protect those whom are among the most vulnerable
    \item assume the burden of proving the output of a tool as the least harmful course of action 
\end{enumerate}

\section{Discussion}

The current paper argues administrative policy not only limits which values technological tools like AI can or should embody, but also predetermines how those values are operationalized within AI products. Inspired by the  value source analysis method posited by \citet{friedman_survey_2017}, we use a computationally enhanced case study of administrative child welfare policy in New York State to identify 13 operationalziations of justice, which we term 'justice principles'. The principles illustrate the nuance required of AI tools designed to facilitate decision-making in the child welfare context specifically. We then offer 8 design guidelines for use in the development of AI tools in human service systems, broadly defined. We argue these guidelines go beyond child welfare systems specifically: they are the design guardrails which should serve as the upper and lower bounds of any AI or other technological tools developed for use in human service systems.  Importantly, we define these guidelines based on our extensive qualitative analysis of policy documents directly. For example, an AI tool designed using these guidelines would define the best outcome as a process whereby primary safe relationships are strengthened and nurtured. Such an AI tool might focus on restorative practices to facilitate the adjudication of grievances in a criminal justice setting; foster a process of group supported decision-making in a health care setting; or emphasize safe, long term kinship care in a foster care setting.

\subsection{How this shapes our understanding of AI in child welfare}

Failures of existing AI solutions in child welfare might, we argue, be viewed in large part as they byproduct of tools which prioritize certain principles of justice at the expense of others, rather than a failure of considering justice at all. Said another way, we wonder if it is possible the failure of existing tools in the child welfare space might be at least partially explained by the over-emphasis of a tool on one principle of justice over another, rather than attempting to balance across the spectrum. For example, Birth match arguably holds as its primary principles protecting from exposure to harm and accountability. However, the tool neglects the other identified values, chief among them the proprietary rights of family, relationships of care, and transparency. In doing so, the tool not only fails families \cite[as has been established in prior work][]{lanier2020preventing}, but also fails to operationalize \emph{all} the principles of the system.

Of course, existing operationalziations of values in policy may not always be the \emph{right} operationalizations. However, solutions that stray too far from, for example, operationalizations of justice espoused in policy are also likely to fail, because they cannot account for the balancing act the system attempts to engage in. Such solutions, often framed as ways of constructing better or more inclusive value systems into AI tools, end up proposing new values, or new operationalizations of existing values, into the system. But such efforts, at least without a coordinated push for policy shifts, is also likely to fail. For example, one proposed solution has been to focus AI implementation through an examination of the latent factors present in families who exit the system without a substantiated case \cite{Rodriguez_DePanfilis_Lanier_2019}, which suggests "some latent factor" might be better understood through AI. Such a solution values transparency, relationships of care, and equity, but more concretely attempts to elevate a new value: family strengths. This value is in direct conflict with the systemic value of protecting from exposure to harm: adopting this proposed solution might require the system to abandon one value for its opposite, a difficult endeavor for even the most flexible of systems. 

\subsection{VSD, Values and Policy}

Although the current manuscript does not seek to extend VSD directly, it is inspired by the assertion in \citet{friedman_survey_2017} that value sources may be examined when designing technological tools. However, other than the case study offered in \citet{friedman_survey_2017}, the current authors can find no use of the method in the current literature. Perhaps this omission is due to the tedious nature of policy analysis -particularly one the engages with policy documents qualitatively, as we have done here. We argue the current manuscript provides a generalizable method for replicating policy analysis in a case study approach, which may in turn help move VSD further along the path of employing value source analysis routinely. To our knowledge, this maybe be one of the first papers analyzing administrative policy as a value source for AI specifically. As such, while its inception is necessarily narrow, this manuscript brings us closer to our goal of building a framework in which  AI tools first account for system values and are explicit about how they are prioritized, operationalized, evaluated, and iterated on. This conclusion is markedly different from other work on values in high priority AI applications, since it argues for contextual understanding before AI tool development. 

\subsection{Towards an AI in human service principle design framework}

Arguably, some work in the fairness in ML community surrounding child welfare has prioritized identifying one or more values, or what we term operationalziations of values (i.e. principles), which should be held centrally by AI developers working in high-risk decision making spaces \cite{saxena2024algorithmic}. The present work illustrates that the elevation of one value or principle over another is perhaps misguided given the ways in which administrative policy operationalizes values. Rather, our results indicate  \textbf{AI tools developed for the child welfare space (and indeed many of the high risk decision spaces found in human services) should account for \emph{all} system values and principles, and be explicit about how they are prioritized, operationalized, evaluated, and iterated on}. This is challenging due to the nature of competing values for which AI needs guardrails to navigate, such as policy. Developing a design framework for AI tool development and implementation in human services will facilitate the identification of those guardrails by offering developers and scholars inclusive, yet flexible, optimization parameters. 

The next step in developing this framework is finding a method whereby we can adjudicate tensions for the ways in which different values and/or principles are operationalized and/or used. For example, fairness may be understood as giving everyone the same thing, while equity might be defined as giving everyone what they need. Both are answers to the question of justice (what is the right thing to do?). Using principles as the framework object, future research should examine how system principles move between these two solutions (among other potential strategies, see for example \cite{thacher_managing_2004}), offering insight into how AI tools can resolve system priorities while balancing system principles.

\section{Conclusion}

Well established is the notion that AI solutions in human service systems like child welfare should privilege a particular way of understanding justice. What is less clear, but equally critical, is an understanding of how these privileged views of justice align with those espoused and/or implied by the system. We present here an  analysis of the way in which one critical value, justice, is operationalized, offering our findings as 13 justice principles. We describe how justice defined in any given AI solution should intersect with the operationalizations embedded in the policy governing the system in which those tools are meant to operate. We conclude our analysis by offering 8 AI design guidelines, for use by researchers and developers building AI solutions for human service spaces.    

Our work has a number of important limitations. First, our analysis is a case study of administrative policy documents in a single U.S. state. The case study approach is justified given the hundreds of hours of qualitative research required to study it, as well as the contextual importance of NYS in both child welfare and policy we noted previously. Scholars should nonetheless take care in extrapolating our findings more broadly, particularly in assuming our operationalizations of justice in NYS align with those espoused in other states.  In addition, our work represents the views of a diverse, but still limited, set of perspectives. Future value source analysis efforts should endeavor to include perspectives from system-impacted individuals. Finally, additional perspectives on justice, and thus additional constraints on AI systems implemented within child welfare in NYS, might be found in relevant case law, policy guidance documents, and in federal policy. These limitations point to interesting and important avenues for future work in continuing to build out a framework that helps clarify value conflicts between AI solutions and social policy. Nonetheless, the present work still provides an important step in characterizing how the principles of a policy system shape and thus require consideration in the development and proposal of new technologies.

\bibliographystyle{SageH}
\bibliography{combined_cleaned}


%
%
%
\clearpage

\appendix

\section{Substantive Themes Used for Analysis}

\begin{table}[H]
    \centering
    \begin{tabular}{lc}
        \toprule
        \textbf{Category} & \textbf{Frequency} \\
        \midrule
        Rules and Regulations & 3216 \\
        Key Providers & 1323 \\
        Court Rules & 1319 \\
        Required Documentation  & 1108 \\
        Money Rules & 492 \\
        Physical Facility Rules & 308 \\
        Responsibilities of Child Protective Services & 173 \\
        Assessment Rules & 101 \\
        Kinship Guardianship Payment Rules & 87 \\
        Rules for Fostering & 69 \\
        Rules for Executing a Surrender & 57 \\
        Service Plan Rules & 49 \\
        Violation of Rules & 43 \\
        Agency Money Rules & 34 \\
        Rules about Religious Exemptions & 21 \\
        Parental Rights Reinstatement & 9 \\
        Removal without court order & 6 \\
        Rules about Voluntariness & 4 \\  
        \bottomrule
    \end{tabular}
    \caption{Summary of Rule Code Themes with Frequency Counts }
    \label{tab:substantive_themes}
\end{table}

Table~\ref{tab:substantive_themes} shows the 15 substantive themes we selected for use in the present work, out of 282 themes identified in a first round of qualitative coding for a larger project studying New York state policy.

\clearpage

\section{Principle Definitions for Sentence Embedding Analysis}

\begin{table}[H]
    \centering
    \begin{tabular}{p{4cm}p{10cm}}
        \toprule
        \textbf{Principle} & \textbf{Phrases used for Embedding} \\
        \midrule
 Transparency & simple, precise, accessible, detailed record, exact, legible, readable, plain language, explicit, direct \\ 
Well-being & harmony, caring, serene, comfortable, peaceful, haven, cover, security, welfare, safeguard \\ 
 Protecting from Exposure to Harm & equitable, impartial, dispassionate, objective \\ 
 Ethics & reasonable, balanced, nonpartisan \\ 
 Right to Life & human rights \\ 
Proprietary Rights of Familial Relations & proprietary right, possession rights,  \\ 
 Relationships of Care & kindred, kinfolk, household, Consanguinity, filation, filial, phylogentic relation, birth, tribe \\ 
 Autonomy of Children & autonomy, agency, self determination, volition, liberal, liberality,  \\ 
Liberty of Faith& human rights, freedom to worship, freedom of worship \\ 
 Accountability & responsible, liable, obliged, beholden, answerable \\ 
 Fairness & suitability, appropriate, contextual, decent \\ 
 Equity & honorable, conscientious, scrupulous, rightness \\
        \bottomrule
    \end{tabular}
    \caption{Summary of Rule Code Themes with Frequency Counts }
    \label{tab:simcse}
\end{table}

Table~\ref{tab:simcse} provides the collection of phrases used to search for excerpts related to each principle.

\clearpage

\section{Additional Details on Data Collection}

One member of the research team, a PhD candidate with knowledge of child welfare policy and experience conducting related research, was responsible for entering Casetext links into a Google spreadsheet next to their associated citations. This process also involved evaluating the potential relevance of sections surrounding those cited by CWIG’s database. For example, there were several instances in which CWIG cited a particular section at the exclusion of other relevant sections within a chapter or article, in which case the research team member copied the link for the entire chapter or article and entered it into the spreadsheet. Whenever it was not feasible to quickly determine the pertinence of surrounding sections (e.g., when section titles were ambiguous or PDFs were too lengthy to review), the researcher flagged the chapters and articles for further investigation. 

Several research team members then worked together to download PDFs from the Casetext website and manually upload them to a shared DropBox folder. As noted above, some items had been flagged for further investigation because it was not immediately apparent whether they were relevant to the project's aims. For these links, the researcher responsible for entering the citations and links into the spreadsheet skimmed the surrounding PDFs for terms directly related to child welfare service provision or that may be applicable to child welfare and therefore cited within child welfare-specific statutes. For instance, sometimes child welfare policies rely on definitions from sections that pertain to children and youth but are codified elsewhere (e.g., in sections on juvenile justice). Whenever the researcher determined that several sections within a given chapter or article were pertinent to the project's aims, the researcher screened in the entire chapter or article for inclusion within the dataset. In other cases, only a few sections were deemed relevant and screened in for inclusion. The other research team members who participated in the data collection process were instructed to download all the files within each link, including statutes that had been repealed or sections that were labeled by Casetext as reserved. However, it is worth noting that we did not adhere to these guidelines throughout the entire data collection period, given that we only decided to include repealed statutes after the researcher responsible for maintaining the spreadsheet had already collected some of the data.

\begin{figure}[H]
    \centering
    \includegraphics[width=.5\textwidth]{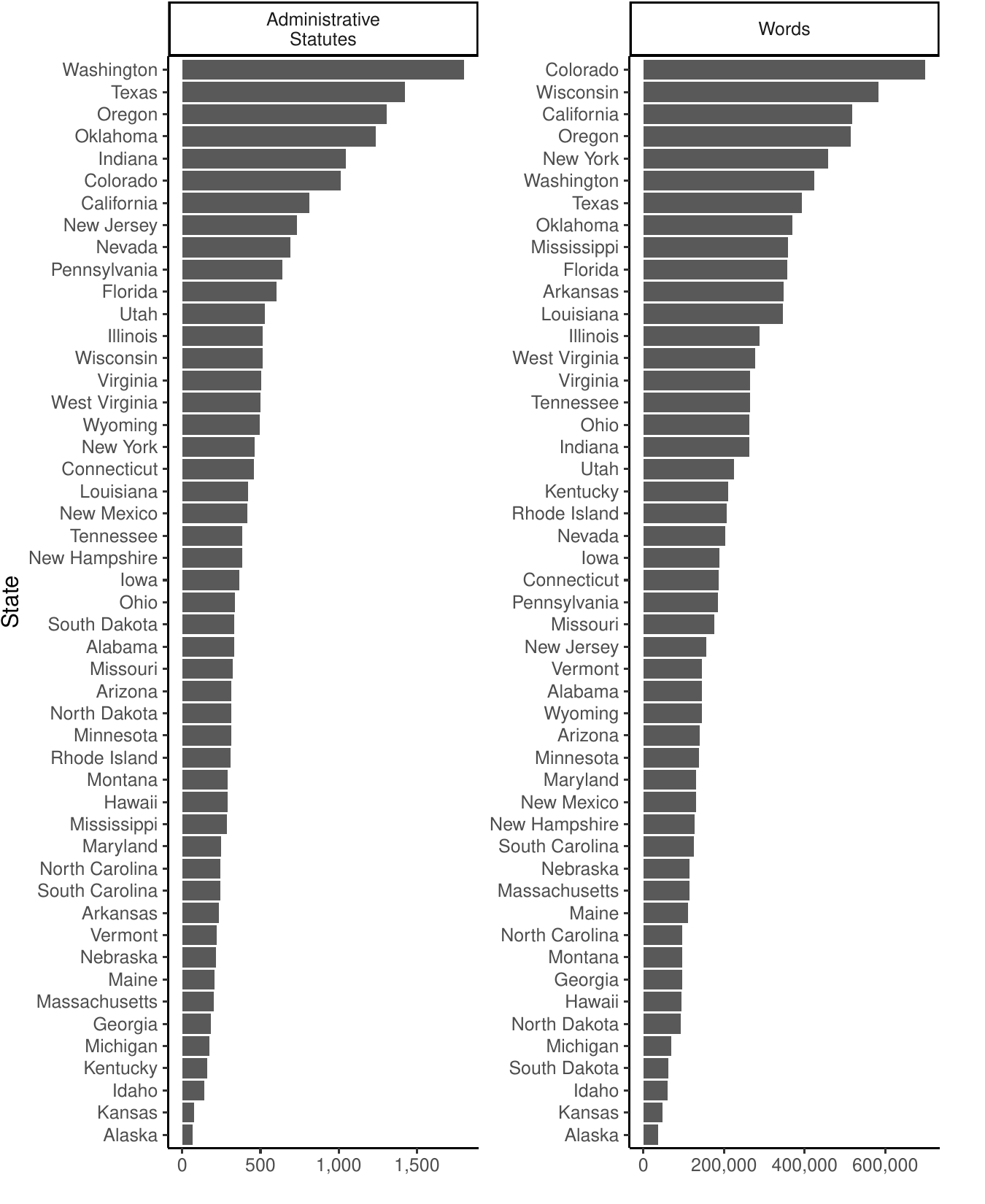}
    \caption{Descriptives of the dataset on a per-state basis. Presented are the total number of words across all state statutes (left subplot) and the total number of administrative statutes  (right). }
    \label{fig:totals}
\end{figure}

Figure~\ref{fig:totals} presents descriptive statistics about the dataset collected. In total, we gathered 23,298 policy documents (i.e., statutes, rules, and regulations) containing over 13 million words, with both the number of child welfare-related policy documents and the overall volume (in terms of word count) varying widely across states. The policy documents of states with larger numbers of documents and/or higher word counts often contained highly specific content. For example, the administrative code for the state of Oregon, which ranked the third in total number of documents, included 17 different sections devoted entirely to outdoor youth programs. However, even some states with smaller volumes covered a high level of detail in their policy documents. One example is the state of Virginia, which ranked 15 across both metrics and still included policy documents related to hunting, as well as the total number of toilets required within child welfare facilities.

These anecdotal points serve to emphasize the volume of child welfare policy collected, and the need for the mixed methods approaches discussed above in our review of the literature too ensure a qualitative coding that was reflective of at least some of the breadth and depth of the content.

\clearpage

\section{Additional Details on the First Round of Coding}

Our first round of coding, relevant to the broader project from which this paper draws, uses a computational grounded theory approach. Specifically, we first apply a topic model as an initial round of coding, which we then use to initialize a codebook for more detailed qualitative coding. In child welfare policy, for example, a ``topic'' might be ``child maltreatment,'' and the words associated with that topic (i.e. the words that have a high value in the posterior probability distribution for that topic over all words) might include ``maltreatment,'' ``harm,'' ``intent,'' and so on.   The goal of a topic model is, given a fixed number of topics, to take in a corpus of documents and to return a set of topics and the topics associated with each document. We use the STM because \citet{rodriguez_computational_2020} find it to be comparable to the first round of qualitative coding, facilitating superficial categorization of text data allowing for in-depth reading of specific texts in the following coding round(s). We apply the STM to policy across all states; following best practices \cite{antoniak2019narrative} we apply basic domain-specific cleaning (e.g. removing policy numbering) and then treat each paragraph of each policy document as a separate document. The final decision is to select an appropriate number of topics. While there is  no scientific consensus on the best way to select the number of topics \cite{grimmerTextDataPromise2013}, \citet{mimno2014low} present an empirically effective heuristic that we make use of here to estimate the number of topics based on a spectral initialization of the model. Though this method provides no statistical guarantees, it is a useful starting point for topic selection. We employ this tool here, and as a result determine a final setting of $k=74$ topics. We estimate the model using the STM package \cite{roberts2019stm}, and once the final model was estimated, we used the STMInsights package in R \cite{stminsights} to manually explore the outputs. The topic model was used to seed the initial qualitative codebook, and is not the focus of the work.  \footnote{All code for the topic modeling will be released upon acceptance of this work.}

\begin{figure}[H]
    \centering
    \includegraphics[width=.5\linewidth]{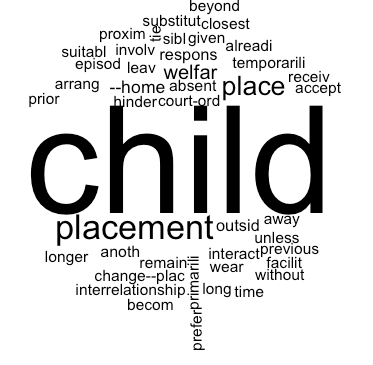}
    \caption{A word cloud representing one topic from our initial topic modeling of all state child welfare policy}
    \label{fig:topic_fig}
\end{figure}

Figure~\ref{fig:topic_fig} displays a wordcloud representing one topic (Topic 73). The topic, and our exploration of relevant excerpts associated with the topic, was used to seed a top-level code in our initial codebook for excerpts pertaining to child placement, with subcodes for placement standards, out-of-home placement, and residential placement.

 \end{document}